\documentclass[12pt]{article} \pdfoutput=1
 \usepackage{amsfonts}
 \usepackage{amssymb,amsmath}
 \usepackage{graphicx} 

 \newcommand{\crlb}[1]{\label{#1}\\[2pt]}
 
 \newcommand{\crld}[1]{\label{#1}}
 \newcommand{\eela}[1]{\quad\hbox{\scriptsize{#1}}\label{#1}\end{eqnarray}}
 \newcommand{\eelb}[1]{\label{#1}\end{eqnarray}}
 
 \newcommand{\newsecb}[2]{\section{#1}\label{#2}\setcounter{equation}{0}}

 \newcommand{\nolabels} {\def\eel{\eelb}\def\eeql{\eeqlb}  \def\crl{\crlb} 
 \def\newsecl{\newsecb}\def\bibiteml{\bibitem} \def\citel{\cite}\def\labell{\crld}}
\newcommand{\eeqla}[1]{\quad\hbox{\scriptsize{#1}}\label{#1}\end{aligned}\end{equation}}
\newcommand{\eeqlb}[1]{\label{#1}\end{aligned}\end{equation}}

\newcommand\publishversion  {\nolabels\setlength{\textheight}{8.3in}\setlength
    {\oddsidemargin}{0in} \setlength{\textwidth}{6.3in}\setlength{\topmargin}{-0.2in}}

\def\beq{\begin{equation}\begin{aligned}}		\def\eeq{\end{aligned}\end{equation}}
\def\be{\begin{eqnarray}}  					\def\ee{\end{eqnarray}}		
   \def\bi#1{\begin{itemize}\item[#1]} 	      	   \def\ei{\end{itemize}} 
   \def\eqn#1{(\ref{#1})}
   	 \def\fn{\footnote}	 

		   	   	\def\del{\delta}        \def\lam{\lambda}  
		              
		\def\gam{\gamma}    		  	    
		\def\m{\mu}	    		\def\f{\phi}        		          
		 	 		     	\def\tht{\theta}

	 \def\LL{{\mathcal L}}         \def\pa{\partial}		\def\ra{\rightarrow}	
	\def\bra{\langle} 		\def\ket{\rangle}

\def\fract#1#2{{\textstyle\frac{#1}{#2}}}	 	 	
\def\ffract#1#2{\raise .2 em\hbox{$\scriptstyle#1\,$}\kern-.34 em/\kern-.34 em\lower .15 em \hbox{$\scriptstyle\,#2$}}
\def\half{\fract12}

\def\bpmatrix{\begin{pmatrix}} 			\def\epmatrix{\end{pmatrix}}
\def\bmatrix{\begin{matrix}} 			\def\ematrix{\end{matrix}} 
\def\bcenter{\begin{center}}			\def\ecenter{\end{center}}
\def\lowerheightfig#1#2#3{\(\raise-#1\hbox{\includegraphics[height=#2]{#3}}\)}
\def\lowerwidthfig#1#2#3{\(\raise-#1\hbox{\includegraphics[width=#2]{#3}}\)}
\def\widthfig#1#2{\includegraphics[width=#1]{#2}}
\def\th{\({}^{\mathrm{th}\ }\)}		 
\def\intt{{\mathrm{int}}}

\def\ol{\overline}   
\def\weglaten#1{}	
 
\publishversion
\begin{document}    

\begin{titlepage}
 \title{Projecting local and global symmetries\\[5pt] to the Planck scale\fn{Dedicated to Prof. Chen Ning Yang at the occasion of his 100\th Birthday.}}
 \author{Gerard 't~Hooft}
\date{\normalsize
Faculty of Science,
Department of Physics\\
Institute for Theoretical Physics\\
Princetonplein 5,
3584 CC Utrecht \\
\underline{The Netherlands} \\[5pt]
e-mail:  g.thooft@uu.nl 
\\ internet: http://www.staff.science.uu.nl/\~{}hooft101 \\[5pt]
January 18, 2922}
 \maketitle
\begin{quotation}  \vskip-20pt
\noindent {\large\textbf{Abstract}}\ 	\\[10pt]
	The Standard Model of the elementary particles is  controlled by more than 20 parameters, of which it is not known today how they can be linked to deeper principles.
Any attempt to clean up this theory, in general results in producing more such parameters rather than less. Yet it is clear that, at distance scales of the order of the Planck length, the gravitational force presents itself in such a way that the need for new physical principles is evident. \\
 A stand-in-the-way is then quantum mechanics, a theory that demands the occurrence of superpositions of physical states in such a way that, when combined with general relativity,  space and time themselves may require new formalisms for being used as primary frames for the descriptions of events.\\
In previous papers the author proposed that quantum mechanics as a theory for the elementary particles should be rephrased as originating from a combination of deterministic evolution laws and discreteness at the Planck scale. 
This may well have a drastic effect on the symmetry structures and algebras. Local, discrete and continuous symmetries do not 
emerge without a cause, and we suggest that the symmetries can tell us more about fundamental constants, among which the Higgs mass is the most peculiar and the most challenging one. 
\end{quotation} 
 \end{titlepage}
	
\newsecl{Introduction}{intro.sec}  \setcounter{page}2
In 1954, Chen Ning Yang and Robert L.~Mills published\cite{YM.ref} their idea that Maxwell's equations for the electro-magnetic force could be generalised, such that the underlying local symmetry involving the vector field \(A_\m(\vec x,t)\) would turn into a non-Abelian one, and the quanta of this vector field would become more complex than the familiar photons, as they would interact directly with one another, and instead of a single type of electric charge, there would be an algebra of charges. 

The properties of the interactions are almost completely determined by this algebra, but an important excieption has always been the universal strengths of these interactions; it has been generally agreed upon that the interaction strengths are given by one freely adjustable parameter for each invariant subgroup of the algebra, forming a set of constants of nature, whose actual values are freely adjustable in the theory, so that they can only be determined by meticulous measurements. 

The discovery of  Yang-Mills fields may be regarded as one of a few elementary milestones of our journey into the realm of elementary particle physics. It inspired other scientists to study the possibility that photons of the Yang-Mills type might exist, hidden somewhere in the world of elementary particles, and in the mean time  investigate various types of currents that may be related to the Yang-Mills currents, and could play a role in other interactions.

There were other milestones in this field, such as the discovery that the Yang-Mills interactions tend to be modified by vacuum effects, leading to the existence of scalar particles that would hide the Yang-Mills symmetry\cite{EnglertBrout.ref,Higgs.ref,Kibbleetal.ref}, the so-called Higgs particles. Subsequently it was discovered that almost all interactions among the elementary particles can eventually be reduced to Yang-Mills interactions. Indeed, while most theories for the interactions of elementary particles had the tendency to generate infinitely strong forces and mass terms, particularly when particles had spin,  the Yang-Mills theory was found to be one of the very few theories that allowed for the interaction strengths and mass terms to be renormalised so that the infinities disappeared, in spite of the presence of spin-one particles, and all this without violating important principles such as locality and unitarity.

A conclusion that should be emphasised, is: \emph{if you believe that the Yang-Mills theory may mean something important in particle physics, you should not take just a piece of it} (like focusing only on the currents) \emph{or mutilate it} (such as adding mass terms that violate the local symmetry), \emph{but one should accept the  theory all the way, without modifying its principles.} We physicists  had to learn that the correct way to generate masses for the vector particles is to invoke the Brout-Englert-Higgs mechanism, which does not affect local gauge invariance. The deeper significance of such lessons will be emphasised when we hit upon another example shortly in this chapter.

The Lagrangian \(\LL\) most often used in elementary particle physics may be written symbolically  like
\be  \LL  = -\sum_{k}\fract {1}{4g_k^2} F^k_{\mu\nu}F^k_{\mu\nu}-\sum_j\ol\psi_j(\gam^\mu (\pa_\mu+i A_\mu)+m_j)\psi_j &\\
	-\ \half ((\pa_\mu+iA_\mu)\phi)^2-V(\phi)-\ol\psi Y(\phi)\psi,&
\ee
where we used some abbreviations to represent the real mathematical structure. Here, the index \(k\) counts the invariant subgroups of the algebra, and the coefficients \(g_k\) are the freely adjustable Yang-Mills coupling coefficients,  while the scalars \(\phi\) come with similar interaction coefficients \(\lam_i\) in the expressions for \(V(\f)\), and their Yukawa interactions with the fermions \(\psi\) come with coefficients \(Y_i\)\ .

Not only may these coupling parameters be chosen arbitrarily, in principle, but they also depend on the momentum- and energy scale of the interactions considered, a dependence controlled by renormalization group equations\cite{rengroup.ref}. These facts actually raise important mathematical questions that can be phrased but have not been properly answered. We now propose that the mathematical basis of these theories has not yet been explored  sufficiently rigorously.  This paper is a plea to investigate alleys for improvement here; as yet physicists' attitude has often been that the theories work just fine up to the extent that we really need them.

It has become largely accepted that perturbation theory is reliable as long as the expansion coefficients, \(g_k,\ \lam_i\) and \(Y_i\) are kept sufficiently small. However, what does `sufficiently small' really mean? Indeed, nature does not always keep them small. Since these coefficients depend on the energy scales of the particles involved, there are occasions where the power expansions in these variables do not converge sufficiently well. One may try to avoid these domains, but it is not well-understood whether this can always be done, and how the entire theory should be redefined if we ask for mathematical rigour everywhere. For instance, in the \(U(1)\) sector of the theory, the expansion parameter \(g_1\) tends to blow up, producing a Landau ghost in the amplitudes in the high-energy/small distance domain, exactly where we would demand mathematical accuracy. \emph{Can our theories be defined with mathematical rigour?} It looks as if this is not the case.\\

We return to this point soon, as we have ideas for improvement.
\newsecl{Quantum Mechanics}{QM.sec}
Elementary particles obey the laws of quantum mechanics. Most authors are convinced that these laws are fundamentally different form any classical system, as has been argued by J.S. Bell\cite{Bell.ref}. Bell considered the time evolution of a quantum system, and asked how we should define the concept of \emph{causality,} to relate states or configurations in the future, to what they may have been in the past.

He thought that statistical configurations in the past must be fundamentally uncorrelated, and that 
the outcome of experimental measurements, as predicted by quantum theory and later checked accurately, show correlations that cannot be explained in any deterministic law or model.  Most of these arguments, however, were verbal and not corroborated by complete sets of equations of motion, in any explicit physical model.  

Present author has been suspicious of such arguments for a long time. And indeed, we recently found models that are fundamentally deterministic, and, most importantly, statistical distributions in the past are strongly affected by states a system may be observed to be in in the future. This is because there is information in the present state, and deterministic equations cannot erase such information, so, in a hidden form, knowledge of the present gives highly entangled information (in the classical sense) concerning the states in the past.\cite{GtH-CA.ref}

The way to phrase what we have found in Ref.~\cite{GtH-CA.ref} is, that quantum mechanics should be regarded as a \emph{vector representation} of all physical states allowed in any kind of physical model. 
This underlying physical model may be completely classical and deterministic in the usual sense. 

In a somewhat different wording, the same conclusion was arrived at by Wetterich\,\cite{Wetterich.ref}, who decribes fermion models  in terms of deterministic evolution laws: the only reason why particles obey probabilistic laws is that the initial states are distributed probabilistically. The evolution laws themselves are classical.

    We emphatically include \emph{superdeterminism}\cite{Hossenf.ref}, which here means that also the behavior of the observer, and the settings chosen in his/her measurement devices cannot be viewed independently from the physical states of all other participating particles.

Indeed, assuming an underlying model to be \emph{completely} deterministic removes most of the `quantum paradoxes' that were thought to be special to quantum mechanics alone. \emph{If one is prepared to consider deterministic theories for elementary particles, one must accept determinism, as well as `superdeterminism', all the way}\cite{Hossenf.ref}.

An important example here is the way physicists assumed \emph{statistics} should be incorporated in a causal, deterministic model. A model may only be truly deterministic, and truly causal, if its dynamical variables do not depend on real numbers but only on integers, or perhaps also some rational numbers. In this case, it will be a lot more straightforward to employ \emph{time reversal}. If real numbers would be involved, the statistical behavior of a system such as a gas or liquid, consisting of atoms, would exhibit chaos, controlled by Lyapunov exponents. Statistical mixing would occur naturally, and assuming statistical independence would then be possible. However, one can easily imagine models where time reversal symmetries hold, but where only integers describe the relevant variables. The number of physically allowed states would then naturally be limited by thermodynamical rules. What happens can be visualised in computer simulations: motion backwards in time may seem to generate statistical mixture, but regardless how far we go to the past, we shall find that statistical mixture is only apparent, the model going backwards in time generates as many correlations, and as much entanglement, as the model forwards in time.

This is particularly of importance in the interpretation of Bell's theorem. If, in his Gedanken experiment, either Alice or Bob `choose a different setting', for instance by switching a variable in their computer simulation, one finds that, at all times, statistical correlations between photons my have altered, in the entire past as much as in the entire future.

The key element in our classical models that overturns Bell's theorem is the observation that \emph{superpositions} are man-made structures that describe probabilities in quantum mechanics, while probabilities in a classical theory do not include quantum superpositions. Forbidding quantum superpositions in the classical description affects the way statistical distributions propagate in time. It implies that, whenever Alice and or Bob make even the slightest modification in their settings of an experiment, quantum physics would assume the photons they measure to be superpositions of the previous photons, while in the classical descriptions these photons are altogether different. Bell assumed that superimposed photons evolve into superimposed photons. Quantum mechanically they do, but classically they can't, while the classical probabilities can be modelled in such a way, that the quantum experiments are indeed reproduced. I refer to our latest papers\cite{GtHFF.ref} where these features are elaborated.

While one might accuse philosophers of science of having not gone deeper into the equations of motion for classical systems,  one might also criticise physicists, in turn, for not taking the problems raised by philosophers seriously enough. Remedies of the putative shortcomings will affect the Standard Model almost certainly when we try to incorporate `quantum gravity' effects. Usually, this is considered to be the domain of superstring theory, and philosophers are asked to get out of the way. But string theory\cite{strings.ref} is not deterministic by construction\fn{But its state counting properties might suggest otherwise, see Ref\cite{GtHstringdet.ref}.}, and we now suspect that this is an essential weakness.

In short, the lesson we had learned with Yang-Mills theory was that local gauge invariance had to be respected as a primary principle  all the way, instead of partly hanging on to older ideas. 
It is not self-evident, but we propose  that here also, we should pick up this lesson right at the present point, by saying that determinism should be embraced all the way as a primary physical principle, that is, we demand it to hold without any exceptions. And now we propose to unravel the consequences of this principle in the case of the Standard Model and the question how it is to be linked with the gravitational force. 

In the past, it was `quantum mechanics' that was thought to be the fundamental basic principle for all our theories about the smallest units of mass, energy, momentum, and so on, but time has come  to accept the discovery that there is no strict dividing line between quantum mechanics and classical logic. We found basic models that can be handled entirely as if they were quantum models, but that can be observed to obey the strictest form of determinism at the same time. These models are also essentially based on \emph{discreteness}, and this may indicate that freely adjustable parameters in theories such as the Standard Model, should actually be limited to discrete subsets. What these subsets exactly are, is still beyond our ability to determine, but we hope that further research here will bring new light.

One of the most serious concerns in our understanding of quantum mechanics, has been the issue of \emph{locality}.
In Bell's experiment, as in many other quantum `paradoxes',  it seemed that `spooky signals' may be travelling faster than light, just so as to restore information that was thought to be lost due to statistical independence. Again, these are mere appearances. The `spooky signals' behave totally normally if we follow them backwards in time. We emphasise again that the models we start off with, are ordinary deterministic models, such as a classical cellular automaton. In terms of the variables in such models, all signals are constrained by speed limits, so that our classical systems obviously do not support  signals going much faster than light. All we now have to add to this is that our quantum systems are based on exactly such models. Spooky signals may arise in our descriptions of the vector representation that reproduces our Schr\"odinger equation, but since all real information of the system is in its classical variables, these spooky signals cannot be used to transmit observable signals. 

Our theory predicts that `real' quantum miracles will never occur; however, this explicitly includes Planck length physics. Therefore there may be quantum effects at the atomic scale that can only be mimicked by classical systems after scaling the latter to Planckian dimensions.
Quantum features more exotic than that must be left to where they belong: in the science fiction literature.

\newsecl{Classical models underlying quantum mechanics}{Classical.sec}

Consider a system that only has a finite number of states (later, one can always consider the infinite limit case). The evolution law dictates how the system evolves in one smallest time step \(\del t\). Such a system must be periodic, but the period may depend on the choice of the initial state. One finds that such a model may always be regarded as a combination of mechanical parts of the form of Figure~\ref{classquant.fig}\(a\). A generic quantum system may be much more complicated, see Fig.~\ref{classquant.fig}\(b\). Nevertheless, as we elaborated in Ref.\cite{GtHFF.ref}, one can approximate any generic quantum system in terms of combinations of the classical form \ref{classquant.fig}\(a\) provided the latter be large enough. This may come as a surprise, and, at first sight, it may sound unbelievable. But then one must remember that today's experiments in physics have access only to a tiny subset of Hilbert space. There are huge numbers of states containing particles that may be heavier than all the ones presently known, or simply contain more energy and momenta than one ever had the chance to observe. The basic argument is summarised as follows.

\begin{figure}[h]
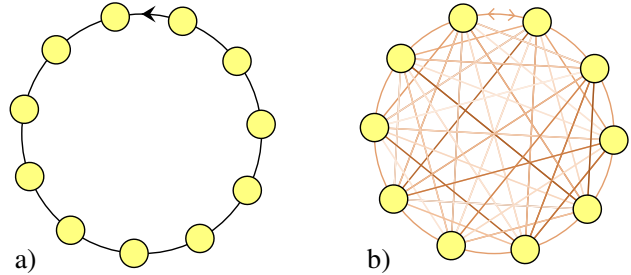
 \bcenter
\widthfig{240pt}{periodiccircle.eps}
\caption{ \small a) Essential part of a classical model. In a bounded universe the model must be periodic. Generic classical models are built by combining ingredients all of the same basic form. b) Generic quantum system. Transitions consist of states that can evolve into superpositions of states. In the text it is explained how a generic quantum model can be obtained by treating \,b)\, as an approximation to much more complicated classical models, even if these essentially only consist of parts of the form  a). \labell{classquant.fig}}
\ecenter\end{figure}

We consider a (quantum) system that allows for all sorts of transitions, as sketched in Fig.~\ref{classquant.fig}\(b\). This means that the Hamiltonian maps every state onto any kind of superposition of states. Notably, this is assumed to be the case even if the basis elements of Hilbert space, the yellow circles in this figure, have been characterised exclusively in a local formalism. In the case of a single particle, this would be the position elements \(|\vec x\,\ket\), but in general we shall be more interested in quantised field theories, where we demand that the basis elements are characterised by the fields, \(\{\phi(\vec x\,)\}\) in position space, where \(\phi\) stands symbolically for all types of (quantised) fields. Clearly, this would include the Standard Model, although we shall have to add states with constraints, invisible under ordinary circumstances, as will be made clear.

Our claim is now that, given any  margin of accuracy,  a classical model can be constructed that will reproduce all probability expressions of the above quantum system, provided that the initial state is one where the local observable is specified, which means that we exclude superimposed states for the initial state. One may then inspect all states that evolve from this initial state and check that our classical model generates the same probability distributions as the quantum system.

At first sight our constraint on the initial state may seem to be too restrictive, but the next assumption is that such a model may define a universe that started as a Big Bang, in which indeed the initial field values were not entangled. This assumption is all we need, to prove our theorem.

But, as the reader may have suspected, there is an other requirement in our construction: the classical model is assumed to include \emph{fast fluctuating variables}. These are variables that evolve into different states within stretches of time short compared to \(\hbar/E\), where \(E\) is the maximal energy that can be delivered to any particle in our quantum model. For example, one may think that, in accordance with most Grand Unified Field Theories, particles will exist with masses way beyond our allowed quantum energies \(E\). Such particles will have to be omitted in our considerations. In practice this means that the fields of these fast fluctuating variables are in their \emph{lowest energy stare}, that is, we only use their vacuum expectation values in our model, but we do not desire to include the actual particles, which are too energetic to be observable in any experiment.

The fast fluctuating variable is an extra ingredient of the form described in
Figure~\ref{classquant.fig}\(a\), which at first sight is completely classical. The classical states evolve periodically, taking all \(N\) values on the circle within the indicated stretch of time. The period is \(T=N\,\del t\). 

Central in our argument will be the concept of energy. The fact that the fast fluctuating variables move faster than we can detect, in practice means that they are assumed to be in an even distribution of all possible states on the circle. To describe this mode with all probabilities equal, we borrow the vector notation from standard quantum mechanics. If the periodicity is \(N\) steps, the evolution operator over time \(T=N\del t\), is the identity operator. Therefore the \(N\) eigen values of the evolution operator \(U(t)\) are \(e^{-i E_n t}\), with \(E_n=2\pi n/T\,,\ T=N\,\del t\). The integer \(n\) takes any value between 0 and \(N\).

To describe the  completely even state, we use the energy ground state, \(E=0\). The higher energy states, \(n=1, \cdots, N-1\), will hardly ever be excited, as in ordinary quantum mechanics. The available energies are too low.

Here we see the emergence of superimposed states. At the stage we are discussing now, superposition is a very simple consequence of the necessary smearing over a stretch of time equal to \(T\). Now comes a crucial step. The fast fluctuating variables will interact, classically, with the variables that describe the (localised and observable) basis states of our model. The latter will be called `slow' variables.

Details of this interaction are described in Ref.~\cite{GtHFF.ref}. The interactions are classical, as are the basis states of our model, but we may use the quantum notation.  Typically, they will take the shape of \emph{exchanges of the slow states, } depending on the positions of the fast variables  (one could hardly imagine other classical interactions). All interactions of our quantum fields 
\(\phi(\vec x\,)\) will be slow, and these interactions can be mimicked by considering a multiple of fast variables, which  are now postulated to dictate exchanges among the slow states only rarely, depending on their precise positions in their respective cycles. They will mostly be found in positions where no interactions take place, so that the interactions that do take place are suppressed when we calculate them:
\be H_\mathrm{slow}=\bra E=0\,|U^\intt |\,E=0\,\ket\ , \eel{perturb.eq}
where now we may use that the state \(|E=0\,\ket\) for the fast moving objects have the effect of averaging out the fast -- but rare -- interactions \(U^\intt \). They are represented by the crossing lines in Figure~\ref{classquant.fig}\(b\).

Our calculation of the effective Hamiltonian for the slow variables in Eq.~\eqn{perturb.eq} is of the standard perturbative type. This is exactly how  Grand Unified fields are handled traditionally; in practice, only the vacuum expectation values of the most massive fields are of any importance, and these expressions are subsequently handled in terms of perturbation expansions.

For further illuminating details we refer to \cite{GtHFF.ref}. The conclusion is that, we should consider  classical models where some variables move very fast while dictating classical transitions among other, slow variables. The fast variables, in turn, are not affected much by the slow ones. As such they therefore violate Newton's action = reaction principle, and this is how we now explain the emergence of  quantum mechanics.

Note that, in precise calculations, the flat distribution of the fast variable states will acquire low amplitude, wave-like correction terms. These are the \emph{virtual} excited particles, just as they occur in Grand Unified Theories. Since the vector notation that we use is exactly valid, these excited states still behave as in ordinary quantum mechanics.

\newsecl{The Standard Model}{SM.sec}

At first sight, the considerations sketched above do not seem to have a great impact on the Standard Model of the subatomic particles and the way we handle it, but, to the contrary, the implications could be sizeable. If indeed our standard theories would be reconsidered as having their roots in a set of deterministic laws of nature, as the present author now believes, there would be at least two important changes in the way we should handle the theory. 

First, local determinism would imply the existence of cellular automaton rules based on evolving data that take the shape of integers only. This would impose severe limitations on the possible constructions one can try to fathom, but the models that will emerge must be denumerable, or possibly finite in number. Consequently, each candidate theory one can construct would be based on relatively simple, rational numbers. In particular, coupling parameters would also not be allowed to range over all possible real numbers, but must be limited to rational numbers, probably limited to be within some  finite intervals. 

Quite a few authors have been speculating that the standard coupling parameters could run towards  different values at different points in space-time. In our scheme, this would be really difficult, well-nigh impossible,  to allow for. If one ever hits upon such features, our theory would be falsified.

Then there is a second observation, which involves the uses of perturbative expansions. The cellular automaton that we believe to underlie the quantum nature of the Standard Model may well allow for perturbative expansions, but there is no reason at all to expect such expansions to converge. The question about convergence of the perturbation expansion is often raised with respect to quantised field theories. Lack of convergence would seem to jeopardise the internal consistency of the model. Now, we emphasise that perturbative field theory itself is an approximation that is no longer needed for \emph{defining} the basis of the theory. Perturbation expansions are as good as they go, but the theory itself will have to be defined in terms of the cellular automaton's (exact) evolution equations. From an axiomatic point of view this would be a great improvement of our understanding.

Unfortunately, there is an important impediment barring us from commencing to implicate our program, which is the gravitational force. At all distance scales where our understanding could be checked, we are confronted with diffeomorphism invariance (general relativity), so one of the first things to do now is to pave our way through that principle. Reconciling cellular automaton rules with diffeomorphism invariance is notoriously difficult. The present work does relieve us from the requirement to do this in terms of Hilbert space, but the cellular automaton must be based on some type of lattice. Diffeomorphism transformations could turn the lattice into a kind of glassy structure. How this should be phrased is not clear at all.

Our philosophy is to advance by one step at the time. The question that must be addressed now is what cellular automaton rules could even approximately lead to the particle spectrum and the interaction algebra known today. Let us start with the symmetries, observed to be exact or approximate features of the Standard Model.

Well-known today is the local symmetry based on \(U(1)\otimes SU(2)\otimes SU(3)\). These are continuous groups, but one can try first to reduce these to a subgroup based on integers. As stated above, integer numbers are mandatory in our theory -- we should not drop them lightly. It may well be that we have to replace the usual local symmetry group by a unitary transformation 
group based on integers. This can be done on any lattice. Vector fields, based on integers, can be defined on the links of the lattice. So far, so good.\fn{Alternatively, one could suspect that the vector representation, introduced above to represent emergent quantum mechanics, might also enhance the symmetries to become continuous transformation groups. The author has not made up his mind as to which route to take here.}

We have fermionic fields in the Standard Model. In discrete theories, these can be very well represented by Boolean variables \(\tht_i(\vec x,t)\in \{0,1\}\)\,. Anti-commuting operators are easy to define there, so we expect Dirac fields to emerge from this, but we do emphasise that not only the Yukawa interaction terms, but also the kinetic ingredients of the Dirac equations can only be ascribed to interactions with fast variables. Presumably we have Boolean fast variables there; this is certainly also allowed.

There is one very remarkable feature of the Standard Model, which is a relatively light Higgs boson (125.3 GeV). In our usual description in terms of quantised fields, this light Higgs particle presents a problem: \emph{renormalisation counter terms play an important role in the theory, and, if extrapolated to the energy scale where the theory must be united with gravity} (the Planck scale), \emph{ these terms are tremendously larger than this mass term, which seems to emerge as balancing in a quite unnatural way.} It is now believed that nature seems to be telling us something important here\cite{Jegerlehner.ref}.

In terms of the cellular automaton, this `unnatural' feature can be rephrased in a meaningful way. Consider the field \(\phi\) of a scalar particle -- such as the Higgs boson. At the Planck scale, let us consider a translation of the type
	\be \phi(\vec x,t) \ra \phi(\vec x,t)+C^{\mathrm{nst}}\ . \eel{HiggsGoldstone.eq} 
Here,  \(\phi\) is the Higgs field, and also the constant \( C^{\mathrm{nst}}\)   stands  for a complex, two component \(SU(2)\) spinor.  We now consider the effect this transformation has on the Hamiltonian of the system. 
The Higgs mass is extremely small at the Planck scale. Furthermore, the effective 4-field self interaction of the Higgs field is small there as well; this is the surprising result of some renormalization group calculations\cite{H.ref}.
Suppose that, in a sense, this interaction can also be considered  to be zero. Then, we find that \emph{the Hamiltonian \(H\) of the theory is almost invariant under the transformation} \eqn{HiggsGoldstone.eq}.

As is well-known, this is Goldstone's principle: the `massless' Higgs particle is the Goldstone particle of a nearly perfect global symmetry.\fn{Note that also the tiny effects of the \(W\) and \(Z\) mass terms violate the Goldstone symmetry.} There is no associated \emph{local} symmetry, other than the usual local gauge symmetry \(SU(2)\otimes U(1)\), because that would be broken by the kinetic term in the Higgs action.
In terms of our cellular automaton interpretation, only the exponentiated Hamiltonian, in the form of the evolution operator, is to be used to describe the evolution. We see that the evolution operator is (almost) invariant under transformations of the type \eqn{HiggsGoldstone.eq}.

This invariance is quite non-trivial. It is not exact, but deviations will be of the order of \(10^{-34}\) or so. This could be an indication that large integers enter in the interactions, integers that play a decisive role in violating this pseudo-Goldstone invariance at such a small rate. One might suspect that, without this approximate global symmetry, not only the Higgs would have been unobservable, but, through the Yukawa interactions, also all leptons and quarks would have masses at the Planck domain.

Eventually, all known matter in the universe may owe its existence to this very tiny global symmetry violation. This is our way of phrasing what Jegerlehner\cite{Jegerlehner.ref} and Veltman\cite{MJGVeltman.ref} have been talking about.

The fact that, when compared to the Planck scale, the leptons and quarks are very light as well, may be due to similar, almost perfectly invariant, transformation rules. As we are dealing with fermions here, the Goldstone fields here will have to be fermionic as well. This may imply that, at the Planck scale, we expect to have very weakly broken supersymmetries, one supersymmetry generator for every fermionic field component in the Standard Model. This symmetry differs from conventional supersymmetry in the sense that it is a global supersymmetry, and the fact that it is not a perfect invariance, giving masses to the particles.

One complication can easily be dealt with: the constant in the transformation rule \eqn{HiggsGoldstone.eq} is not invariant unde the local gauge transformations of the Standard Model; this is taken care of by splitting the Higgs field variable into a unimodular part, \(u^i\) and the modulus \(r\):
\be \f^i(x)=r(x)\,u^i(x)\ , \ee
so that the Goldstone symmetry operators only act on \(r\) and the local \(SU(2)\) symmetry only acts on \(u^i\); this ensures that the Goldstone symmetry commutes with the local \(SU(2)\) transformtions.

In terms of cellular automaton interactions, the fermionic global supersymmetries will be manifest as Boolean symmetries, symmetries that square into the identity operation; by itself, this feature would not be problematic.

We are still at a pioneering stage of our general insights. What is missing is the general equivalents of the Hamilton and Lagrange formalisms, which have been auspicious for understanding quantised systems and their relations with the classical limit, \(\hbar \ra 0\). We now need to know how this works when special types of quantum systems are mapped to deterministic classical systems.

 \end{document}